\documentclass[11pt]{article} 
\usepackage{amssymb} 
\usepackage{amsmath,amsthm} 
\usepackage{graphicx} 
\usepackage{amsgen, amsfonts, amsbsy} 
\usepackage{mathrsfs} 
\usepackage{color}

\newcommand{\be}{\begin{equation}} 
\newcommand{\ee}{\end{equation}} 
\newcommand{\bc}{\begin{center}} 
\newcommand{\ec}{\end{center}}

\begin{document} 

\def\theequation{\arabic{section}.\arabic{equation}} \begin{titlepage}

\title{A simplified climate model and  maximum entropy production}

\author{Valerio Faraoni \\ \\ 
{\small Department of Physics \& Astronomy, Bishop's 
University}\\ 
{\small 2600 College St., Sherbrooke, Qu\'{e}bec, Canada 
J1M~1Z7} 
} 
\date{} \maketitle 
\thispagestyle{empty} 
\vspace*{1truecm} 

\begin{abstract}

A simplified climate model based on maximum entropy production, described 
by a variational principle, is revisited and an analytical solution to its 
Euler-Lagrange equation is found. Mindful of controversy about maximum or 
minimum entropy production in open thermodynamical systems, we show that 
the solution extremizing the action integral corresponds to a maximum.

\end{abstract}

\vspace*{1truecm}
\end{titlepage} 

\clearpage
\bigskip \small

\section{Introduction}
\label{sec:1}

The principle that open thermodynamical systems in nature tend to maximize 
entropy production (MEP) has been debated extensively in the earth 
sciences, and in atmospheric science in particular. The climate is an open 
thermodynamical system in which energy flows from tropical to polar 
regions. A simple model based on MEP was proposed by Paltridge in 1975 
\cite{Paltridge75} to calculate the temperature $T(\vartheta)$ and the 
cloud 
cover as functions of the latitude $\vartheta$. 

MEP has been the subject of much discussion and controversy 
\cite{Goody2007,Caldeira2007,NicolisNicolis2010}. Studies aimed at testing 
this model use simple one-dimensional energy balance models 
\cite{Paltridge75,Paltridge81,Paltridge81,Grassl81,NodaTokioka83,Lorenzetal2001,PujolFort2002,Pujol2003,Kleidon2004,Kleidon2009,Kleidon2010,JuppCox2010,Herbertetal2011a}, with a few exceptions studying  
general circulation models 
\cite{Mobbs1982,Kleidonetal2003,Kleidonetal2006,Kunzetal2008,Pascaleetal2011b}.  
Although, in numerical studies of these models, one can check easily 
whether entropy production 
is maximized, minimized, or has  a saddle point, in these studies one is 
always committed to particular choices of parameters and initial 
conditions, and numerical confirmation does not constitute mathematical 
proof. Some authors in the earth sciences contend that that there is no 
need for  a general mathematical proof and that numerical checks are 
sufficient, but we argue that this attitude is contrary to the spirit of 
science. Anyway, it is not difficult to obtain a proof for simple systems, 
as shown below.   From the mathematical 
point of view, most statements about MEP are not backed by rigorous 
mathematical proofs \cite{Dewar2005,GrinsteinLinsker2007} and this 
probably reflects the fact that MEP is not understood. It is quite possible 
that MEP 
is not a fundamental law, but just an approximation in which different 
scales are separated, but the 
parameter, or ratio of variables, characterizing the MEP regime has not 
been identified. In this case, MEP could be a quasi-static approximation  
in which effects occurring on a longer timescale are neglected in favour 
of processes occurring on a shorter timescale \cite{VincentLabarre}. 

If one takes  a slighly broader view of open non-equilibrium 
thermodynamical systems beyond atmospheric 
models, the spectrum of possible situations appears rich and complicated. 
In one-dimensional diffusion problems for systems in steady state, a maximum 
of entropy production is usually associated with closed boundary 
conditions and a minimum with open ones \cite{Li} (diffusion is 
quite distinct from convection and turbulence, but this example shows that 
opposite outcomes can sometimes occur in the same kind of physical 
processes). 
For other systems, the situation is not clear. Therefore, {\em a priori} 
statements should be backed by testing. A similar controversy on entropy 
production rate can be found in the determination of equilibrium beach 
profiles 
in oceanography. In the zone seaward from the wave breaking point, 
wave friction against the sea bed and transport of sediments dissipate 
energy and a one-dimensional model of this open thermodynamical system 
expressed by a variational principle can be constructed
  \cite{Larson,JI06}, with different authors disputing whether the entropy 
production 
rate is maximized or minimized (recent work determines that it is a 
minimum \cite{Maldonado,MaldonadoUchasara,myJO}). A similar problem occurs 
in the erosion of glacial valleys, where friction\footnote{Friction, 
however, does not coincide with entropy production rate.} is maximized 
instead 
\cite{myJOG2020}, which has also been the subject of a minimum/maximum  
controversy 
\cite{Harbor90,HiranoAniya88,HiranoAniya90,HiranoAniya05,Morgan05}. The  
uncertainty in the theoretical foundations of MEP reflects our incomplete 
knowledge of non-equilibrium thermodynamics. A practical lesson 
gained from the 
literature is that intuition 
often fails in MEP-based 
models and  the nature of the extremum attained by the system should be 
assessed---which is not hard to do numerically---and proved rigorously  
in general whenever possible. 

Here we revisit a simple one-dimensional model with energy transport in 
the meridional direction, based on MEP, which was proposed in \cite{MK}. 
Given the latitudinal distribution of energy absorbed at short 
wavelengths, the model calculates the latitudinal distribution of energy 
emitted at long wavelengths and the meridional heat transport by means of 
a variational principle that extremizes the entropy production rate. The 
Euler-Lagrange equation produced in this way was solved numerically in 
\cite{MK}.

First, we solve analytically the central equation of the model. Second, we 
show explicitly that this solution of the Euler-Lagrange equation indeed 
corresponds to a maximum of entropy production. This is necessary because 
Ref.~\cite{MK} does not prove mathematically that the extremum of the 
entropy production rate is a maximum (a similar, but long and rather 
involved proof for Paltridge's 1975 model \cite{Paltridge75} was sketched 
only twenty years later \cite{ObrienStephens95}).  While the derivation of 
the model's main equation requires only that the solution be an extremum of 
the action integral, to understand the physics and validate MEP it is 
crucial to determine whether this extremum is a maximum or a minimum. A 
recurrent puzzle in MEP-based system is that sometimes there is a maximum 
and sometimes a minimum of entropy production 
\cite{Paltridge75,Paltridge81,Grassl81,NodaTokioka83,Lorenzetal2001,PujolFort2002,Pujol2003,Kleidon2004,Kleidon2009,Kleidon2010,JuppCox2010,Herbertetal2011a,Mobbs1982,Kleidonetal2003,Kleidonetal2006,Kunzetal2008,Pascaleetal2011b,Maldonado,MaldonadoUchasara,myJO}: 
here we {\em prove} (as opposed to checking numerically for special 
configurations) that a maximum always occurs in this model.

\section{The model}
\label{Sec:2}
 
The Murakami and Kitoh one-dimensional climate model in the meridional 
direction \cite{MK} is based on a very idealized radiative  formulation. 
It assumes that:

\begin{itemize}

\item the system is in steady state;

\item the maximum entropy production hypothesis (MEP);

\item the distribution of the absorbed solar radiation 
is a given (even) function $  I(\vartheta)$.

\end{itemize}

The model calculates the long-wave radiation emitted $O(\vartheta)$. 
No assumptions are made about the relationship between heat transport and 
temperature gradient. This model neglects the vertical structure of the 
atmosphere,\footnote{The radiative budget is usually non-local, 
{\em 
i.e.}, a functional of the vertical temperature profile, even in simple 
models when several layers of atmosphere are considered.} cloud radiative 
processes, the oceans, and the atmosphere–ocean coupling.  In reality, the 
incoming radiation $I( \vartheta)$ is not an even function on Earth, 
because of the asymmetry of the albedo, and it does not vanish at the 
poles because of the Earth obliquity.

More in detail: assume that the climate (an open thermodynamical 
system) is in steady  state with energy flowing from the equator to polar 
regions, maximum  entropy production, and that the radiative flux density 
absorbed by the 
Earth at short wavelengths (the insolation), $I(\vartheta)$, is a given 
function of the 
latitude 
$\vartheta 
 \in \left[ -\pi/2, \pi/2 \right]$. 
By symmetry,\footnote{In the end, Ref.~\cite{MK} assumes 
$I(\vartheta)= \beta- \alpha \sin^2 \vartheta$, with $\alpha$ and 
$\beta$ constants.} it must be an even function, 
$I(-\vartheta)=I(\vartheta)$ 
and it must vanish at 
the poles,  $I\left( \pm \pi/2\right)=0$ and decrease going 
from the 
equator to a pole, $dI/d\vartheta > 0$ if $ -\pi/2 <\vartheta<0$ and 
$dI/d\vartheta 
< 0$ if $ 0 <\vartheta< \pi/2 $. Let $ O(\vartheta)=\sigma T^4(\vartheta)  
$
be the flux density radiated by the Earth into 
space at the latitude $\vartheta$, where $\sigma$ is the Stefan-Boltzmann 
constant and $T(\vartheta)$ is the absolute temperature at the same 
latitude. At 
low latitudes absorption dominates and $O(\vartheta) < I (\vartheta) $, 
while 
at high latitudes it is $O(\vartheta)>I (\vartheta)$. 

The net radiative flux density is $ I(\vartheta)-O(\vartheta)$ and the heat 
flux 
transported to higher latitudes is
\be
2\pi R^2 \int_{-\pi/2}^{\vartheta} d\bar{\vartheta} \, \cos\bar{\vartheta} 
\left[ 
I(\bar{\vartheta})-O(\bar{\vartheta}) \right] \,,
\ee
where $R$ is the Earth's radius. The elementary entropy production rate is 
related to the elementary heat production rate $d\dot{q}$ by $d\dot{s} = 
d\dot{q}/T$ and the finite entropy production rate is 
\be
A= 2\pi R^2 \int_{-\pi/2}^{\pi/2} d\vartheta \,  \cos \vartheta \,   
\frac{ \left[ I(\vartheta)-O(\vartheta) \right]}{T(\vartheta)}  
\,,\label{action}
\ee
a functional of the function $O(\vartheta)$. 
The variational principle consists of extremizing this entropy production 
rate subject to the constraint 
\be
\int_{-\pi/2}^{\pi/2} d\vartheta \,  \cos \vartheta  
\left[ I(\vartheta)-O(\vartheta)\right] =0 
\ee
expressing the fact that the climate system is in steady state 
\cite{Rodgers76,MK}.  This constrained variational principle is simplified 
as 
follows in Ref.~\cite{MK}: define $x\equiv \sin\vartheta$ and
\be
y(x) \equiv \int_{-1}^x d\bar{x} \left[ I(\bar{x})-O(\bar{x}) \right] \,;
\ee
then 
\be
y'(x)=I(x)-O(x) \label{y'}
\ee
 (where a prime denotes differentiation with 
respect to $x$) and $y(\pm 1)=0$ \cite{MK}. The action 
integral~(\ref{action}) (divided by the irrelevant constant $2\pi R^2 
\sigma^{1/4}$) is 
converted into \cite{MK}
\be
J\left[ y(x)\right] = - \int_{-1}^{+1} dx \, y'(x) \left[ I(x)-O(x) 
\right]^{-1/4} \equiv \int_{-1}^{+1} dx \, L\left( y'(x), x\right) 
\label{action2}
\ee
where $I\geq y'$ (equivalent to $O\geq 0$) is always satisfied. 
Now we have an unconstrained variational principle $\delta J=0$ with fixed 
boundaries.  Since the Lagrangian $L$ does  not depend explicitly on $y$, 
the Euler-Lagrange equation
\be
\frac{d}{dx} \left( \frac{\partial L}{\partial y'} \right) -\frac{\partial 
L}{\partial y}=0
\ee
gives conservation of the momentum $\Pi_y=\partial 
L/\partial y'$ canonically conjugated to $y$, or
\be
\left( I-y'\right)^{5/4} +\frac{3y'}{C} =\frac{4I}{C} 
\,,\label{1stintegral}
\ee
where $C$ is an integration constant (which is fixed by the boundary 
conditions, as described below). Equation~(\ref{y'}) gives 
\be
C=\frac{ I+3O}{ O^{5/4} } >0 \,.\label{Conce}
\ee 
The first integral~(\ref{1stintegral}) of the Euler-Lagrange 
equation apparently was 
missed in \cite{MK}, where the authors report the second order 
Euler-Lagrange equation, which they integrate numerically.

A second integration is unnecessary. In fact, the 
derivative $y'$ was introduced in \cite{MK} to simplify the 
original variational problem, which 
is solved by determining $O$ or $y'$.
Equation~(\ref{y'}) gives 
immediately  the analytical solution $O(x)$ of the problem through 
\be
I(O)= C O^{5/4} -3O  \label{result}
\ee
or, equivalently,
\be 
I(\vartheta)= C\sigma^{5/4} T^5(\vartheta) - 3\sigma T^4(\vartheta) \,.
\ee
Equation~(\ref{result}) cannot be inverted to obtain $O(I)$ 
explicitly, but this is not crucial.

The integration constant $C$ is fixed by the boundary condition $y(1)=0$ 
(the other boundary condition $y(-1)=0$ is satisfied by construction and 
does not provide new information):
\be
y(1) = \int_{-1}^{+1}dx \left[ I(x)-O(x) \right]=0 
\ee
becomes, using Eq.~(\ref{result}), 
\be
C= \frac{ 4 \int_{-1}^{+1}dx \, O(x)}{
\int_{-1}^{+1}dx \, O^{5/4}(x) } = \frac{ 4 \int_{-1}^{+1}dx \, I(x)}{
\int_{-1}^{+1}dx \, O^{5/4}(x) } \,.
\ee
Alternatively, using the information that $I$ vanishes at the poles, 
Eq.~(\ref{result}) yields
\be
C= \frac{3}{ O^{1/4}( \pm \pi/2) }
\ee
(this value can be obtained also by setting $I(\pm \pi/2) =0$ in 
Eq.~(\ref{Conce})).


\section{Maximum or minimum?}
\label{sec:3}

Consider varied paths $y(x,a)$, parametrized by the parameter $a$, around 
the actual solution $y(x,0)$ that extremizes the 
functional $J\left[ y(x) \right]$, 
\be
y(x,a)=y(x,0)+a \eta(x) \,.
\ee
We have 
\be
\frac{\partial y}{\partial a} =\eta \,, \;\;\;\;\;\;\;\;
\frac{\partial y'}{\partial a} =\frac{d\eta}{dx} \,,\;\;\;\;\;\;\;\;
\frac{\partial^2 y}{\partial a^2} =
\frac{\partial^2 y'}{\partial a^2} =0 \,.
\ee
The second variation of the functional $J$ gives ({\em 
e.g.}, \cite{ArfkenWeber})
\be
\frac{\partial^2 J}{\partial a^2} = \int_{-1}^{+1} dx \left[ 
\frac{\partial^2 L}{\partial y'^2} \, \left( \frac{d\eta}{dx} \right)^2 +
2\, \frac{\partial^2 L}{\partial y\partial y'} \,\eta \, \frac{d\eta}{dx} 
+ \frac{\partial^2 L}{\partial y^2} \,\eta^2 \right]\,.
\ee
Since the Lagrangian $L$  does not depend explicitly on $y$, this 
integral reduces to 
\be
\frac{\partial^2 J}{\partial a^2} = \int_{-1}^{+1} dx \, 
\frac{\partial^2 L}{\partial y'^2} \, \left( \frac{d\eta}{dx} \right)^2 
= -\frac{1}{4} \int_{-1}^{+1} dx \, \frac{ 
\left(I-3y'/8 \right)}{\left( I-y'\right)^{9/4}} \left( 
\frac{d\eta}{dx}\right)^2 \,, \label{integral}
\ee
where $I-y'=O=\sigma T^4 >0 $ so that $I>y'>3y'/8 $, hence the numerator of 
the fraction in the integral 
is positive; the denominator is positive, and $\left( d\eta/dx\right)^2 
\geq 0 $ cannot vanish everywhere. As a result, the 
integral~(\ref{integral}) is 
positive-definite, not only on-shell ({\em i.e.}, when evaluated on the 
extremizing trajectories) but always. Then $\partial^2 J/\partial 
a^2<0$, and 
the extremum is  a maximum \cite{ArfkenWeber} of the entropy production 
rate described by the action integral~(\ref{action2}).

\section{Conclusions}
\label{sec:4}

Neglecting the vertical structure of the atmosphere, the obliquity of the 
earth, the albedo asymmetry between the two hemispheres, and many other 
factors no doubt oversimplifies the real description of the Earth's 
climate, but conceptual box models still have value. They provide insight 
into the essentials of the energy balance without the burden of a myriad 
of complicated details; they can be used for quick tests of numerical 
codes; and they are valuable pedagogical tools.
  
We have revisited the simplified climate model of \cite{MK} and have 
simplified and solved its main equation, which is derived from an action 
integral corresponding to the entropy production rate and involves 
radiative absorbed and radiated fluxes and poleward transport of energy 
from tropical regions. {\em Per se}, the analytical 
equation~(\ref{result}) does not require numerical integration (although 
some numerics are still needed to plot the physical quantities, which are 
given in \cite{MK} following full numerical integration and are not 
reproduced here). Although MEP is routinely verified numerically in this 
kind of model, it is not understood. A first step consists of proving 
rigorously that entropy production is maximum for all values of the 
parameters and initial conditions in their physical ranges.  Moreover, 
there have often been surprises in natural processes associated with open 
thermodynamical systems, in which entropy production is sometimes 
maximized and other times minimized, and it is useful to set MEP on a firm 
footing with rigorous statements before proceeding with numerical studies. 
Indeed, MEP is taken as an assumption or principle, while it is likely to 
be just an approximation valid in a certain regime that is not even 
beginning to being characterized in terms of known variables and 
parameters. Here we have shown that, in the one-dimensional climate model 
of \cite{MK}, the entropy production rate is indeed maximized. Our proof 
could perhaps stand as an example for more realistic models in which the 
atmosphere is stratified and the temperature and the radiative budget are 
non-local. These would be necessary steps in order to make the 
model more realistic.\\\\

{\small \noindent We thank a referee for suggestions leading to 
improvements in the manuscript. This work is supported by the Natural 
Sciences \& Engineering Research Council of Canada (Grant No. 2016-03803) 
and by Bishop’s University.}   
\end{document}